# Identification of a Large Amount of Excess Fe in Superconducting Single-Layer FeSe/SrTiO$_3$ Films

Yong Hu[1,2,#], Yu Xu[1,2,#], Qingyan Wang[1,*], Lin Zhao[1], Shaolong He[1], Jianwei Huang[1,2], Cong Li[1,2], Guodong Liu[1] and X. J. Zhou[1,2,3,*]

[1]National Lab for Superconductivity, Beijing National Laboratory for Condensed Matter Physics, Institute of Physics, Chinese Academy of Sciences, Beijing 100190, China

[2]University of Chinese Academy of Sciences, Beijing 100049, China 3Collaborative Innovation Center of Quantum Matter, Beijing 100871, China

[#]These authors contributed equally to the present work.

[*]Corresponding author: qingyanwang@iphy.ac.cn, XJZhou@iphy.ac.cn

**The single-layer FeSe films grown on SrTiO$_3$ (STO) substrates have attracted much attention because of its record high superconducting critical temperature ($T_C$). It is usually believed that the composition of the epitaxially grown single-layer FeSe/STO films is stoichiometric, i.e., the ratio of Fe and Se is 1:1. Here we report the identification of a large amount of excess Fe in the superconducting single-layer FeSe/STO films. By depositing Se onto the superconducting single- layer FeSe/STO films, we find by *in situ* scanning tunneling microscopy (STM) the formation of the second-layer FeSe islands on the top of the first layer during the annealing process at a surprisingly low temperature (~150°C) which is much lower than the usual growth temperature (~490°C). This observation is used to detect excess Fe and estimate its quantity in the single-layer FeSe/STO films. The amount of excess Fe detected is at least 20% that is surprisingly high for the superconducting single-layer FeSe/STO films. The discovery of such a large amount of excess Fe should be taken into account in understanding the high-$T_C$ superconductivity and points to a likely route to further enhance $T_C$ in the superconducting single-layer FeSe/STO films.**

## I. INTRODUCTION

The discovery of high-temperature superconductivity in single-layer FeSe films epitaxially grown on SrTiO$_3$ (STO) substrates continues to attract intense attention [1–12]. First, they show record high superconducting transition temperature $T_C$ higher than 65 $K$ [1,3,4,7] in iron-based superconductors. This is distinct from the conventional superconductors where $T_C$ usually decreases with reducing thickness [13], even becoming nonsuperconducting when they come to single layers. Second, the enhancement of superconductivity in this system is closely related to the interface effect that provides an exciting way to realize high-temperature superconductivity or novel electronic states in two-dimensional systems. Third, superconducting single-layer FeSe/STO films host a unique Fermi surface topology consisting only of electronlike pockets near the zone corner without hole-like pockets near the zone center [2–5]. This provides an ideal platform for studying the high-temperature superconductivity mechanism of iron- based superconductors. The electron doping mechanism, the interface effect and the origin of high-$T_C$ superconductivity in single-layer FeSe/STO films remain hotly debated [8,10–12].

It has been found that superconductivity in bulk FeSe is extremely sensitive to its stoichiometry; superconductivity can be completely suppressed by the presence of a tiny amount of excess Fe above δ = 0.03 in bulk Fe$_{1+δ}$Se [14]. It is generally believed that the composition of FeSe/STO films grown by molecular beam epitaxy (MBE) is stoichiometric; that is, the ratio of Fe to Se is 1:1 [1,15,16]. Determining the composition and the crystal structure is crucial for understanding the origin of high-temperature superconductivity in single-layer FeSe/STO films. In this work, we report the discovery of a large amount of excess Fe in superconducting single-layer FeSe/STO films. By depositing Se on the surface of single-layer FeSe/STO films, we find the formation of second-layer FeSe islands on top of the first layer during the annealing process by in situ scanning tunneling microscopy (STM). The formation of the second-layer FeSe occurs at a surprisingly low temperature (~150 °C) which is much lower than the usual growth temperature (~490 °C). This provides us a reliable way to detect the presence of excess Fe in FeSe/STO films. We reveal the existence of a large amount of excess Fe in superconducting single-layer FeSe/STO films; the amount is unexpectedly high, well above 20%. This finding should be taken into consideration when determining how superconductivity can survive in FeSe films with such a large amount of excess Fe and what role this excess Fe may play in giving rise to the high-temperature superconductivity in single-layer FeSe/STO films. It also points to a likely route to further enhance $T_C$ in single-layer FeSe/STO films.

## II. EXPERIMENT

FeSe films with different thicknesses [0.45, 0.75, 1, and 1.5 monolayers (ML)] in this study were grown by MBE on high-quality single-crystal 0.5 *wt* % Nb-doped SrTiO$_3$ substrates (Shinkosha STEP substrates), similar to a procedure reported before [1]. The base pressure of the MBE chamber is $1 \times 10^{-10}$ *mbar*. In order to obtain an atomically smooth surface, the STO substrates were degassed at 600 °C for several hours, followed by another annealing at 980 °C for 1 h. The atomically flat terraces were clearly observed in the annealed STO substrates by STM. Ultrahigh-purity selenium (99.9999%) and iron (99.995%) were coevaporated onto the STO substrate from two Knudsen cells with a flux ratio of ~10:1. The STO substrate temperature was held at 490 °C during growth as measured by the pyrometer, and the growth rate was ~0.22 *ML/min*. To obtain a superconducting single-layer FeSe film, the as-grown sample was post-annealed at 530 °C for 4 h. Amorphous Se (~100 *nm* thick) was deposited onto the FeSe/STO films at room temperature in the MBE chamber. The Se-coated FeSe films were subsequently annealed in the MBE chamber at various temperatures ranging from 150 °C to 530 °C. The surface morphology of the FeSe/STO films was examined by the in situ STM (Specs STM Aarhus 150), and all STM images were taken at room temperature. High-resolution angle-resolved photoemission measurements were carried out on our laboratory system equipped with a Scienta Omicron DA30 electron energy analyzer [2,3]. We used a helium discharge lamp as a light source that can provide photon energies of *hv* = 21.218 *eV* (He I). The energy resolution was set to 10 *meV* for the Fermi-surface mapping and band-structure measurements and to 4 *meV* for the superconducting gap measurements. The angular resolution was ~0.3°. The Fermi level was referenced by measuring on a clean polycrystalline gold that was electrically connected to the sample. The sample was measured in vacuum with a base pressure better than $5 \times 10^{-11}$ *mbar*.

**III. RESULTS AND DISCUSSION**

We first start with the preparation of single-layer FeSe/STO films. Figure 1 shows the preparation procedure to obtain single-layer FeSe films on STO substrates, as schematically shown in Fig. 1(a). The STO substrate we used has atomically flat terraces [Fig. 1(b)]. The as-grown single-layer FeSe film has some second-layer patches near the steps, as marked by an arrow in Fig. 1(c). After post-annealing at 530 °C for 4 *h*, the second-layer FeSe patches decompose, and a perfect single-layer FeSe/STO film is obtained [Fig. 1(d)]. The single- layer FeSe/STO film is atomically flat with regular terraces following the STO substrate and is homogeneous over a large scale. Domain structures in the zoomed-in image [Fig. 1(e)] and atomically resolved STM image [Fig. 1(f)] can be clearly observed and are similar to those reported before [17].

We also carried out angle-resolved photoemission spectroscopy measurements on single-layer FeSe/STO films [Fig. 2(a)] which were prepared using the same procedure as that in Fig. 1(d). The Fermi surface mapping of the single-layer FeSe/STO film is shown Fig. 2(b). The carrier concentration is ~0.116 electron/Fe atom, which is determined by the measured area of the electron Fermi pockets. Figure 2(c) shows the band structure measured at 20 *K* along the Fermi surface around the M point. The symmetrized photoemission spectra (energy distribution curves) measured at different temperatures on the Fermi momentum $k_R$ are shown in Fig. 2(d). One can see a sharp coherence peak and a clear gap opening at low temperature. The superconducting gap is ~15 *meV* at 20 *K* and gets closed near 60 *K*. The Fermi surface, band structure, and superconducting gap measured on the single-layer FeSe/STO film are similar to those reported before [2–4]. These results indicate that the single-layer FeSe/STO films we prepared in Figs. 2(a) and 1(d) are superconducting, with a $T_C$ around 60 *K*.

Next we deposited Se on top of the perfect single-layer FeSe/STO film [Fig. 1(d)] at room temperature and annealed it under different conditions to investigate the evolution of the surface morphology by *in situ* STM measurements [Fig. 3(a)]. The annealing sequences are shown in Fig. 3(b), which represents eight consecutive conditions of different temperatures and different times. After each annealing sequence, *in situ* STM image scanning was carried out on the film. For the perfect single-layer FeSe film [Fig. 1(d)] capped with amorphous Se, after the first annealing at 250 °C for 1.5 *h*, a single-layer FeSe film shows up that is covered by a large number of bright islands [Fig. 3(a1)]. With further annealing at elevated temperatures and prolonged time, these islands grow in size, and the number of islands decreases. The total coverage of the islands also decreases [Figs. 3(a2)–3(a5)]; they eventually completely disappear, and the sample returns to the original perfect single-layer FeSe film [Fig. 3(a6)]. To quantitatively determine the coverage of the islands, we sum up the area of all the islands to get the ratio relative to the total field of view. Right after the first annealing sequence, the coverage of the islands [Fig. 3(a1)] is about 23%. The coverage decreases with further annealing and eventually becomes zero [Fig. 3(c)].

After getting the perfect single-layer FeSe/STO [Fig. 3(a6)], we repeated the process and deposited Se on its surface at room temperature again. After the initial annealing at 150 °C for 1.5 *h*, the surface is mostly covered by some disordered structure, and it is hard to follow the morphology of the single-layer FeSe, as shown in Fig. 3(a7). After further annealing at 150 °C for 2 *h*, the single-layer FeSe film fully shows up and is covered by a large number of bright islands [Fig. 3(a8)]. The coverage of the islands [Fig. 3(a8)] is about 23%, very close to the case for Fig. 3(a1). This indicates that the above annealing evolution is reversible. We repeat similar procedures on another single-layer FeSe/STO film, and the results are highly reproducible (see the Supplemental Material [18]). We note that a similar procedure of Se deposition and annealing was used to protect the sample surface and measure the electronic structure and superconducting gap of the single-layer FeSe/STO films [2,3]. A distinct electronic structure and signature of superconductivity at about 65 *K* were observed on the perfect single-layer FeSe films that are identical to the ones in Fig. 1(d) and Fig. 3(a6). These results are also consistent with other measurements [4,5].

In surface science, deposition of amorphous Se onto the prepared films is a common method to protect the surface from being exposed to air and getting degraded [19–21]. This is because amorphous Se can be removed by annealing at a relatively low temperature, usually below 200 °C [21]. It is therefore quite surprising to observe second-layer islands on single-layer FeSe/STO films after annealing at high temperatures above 200 °C, even up to 450 °C [Fig. 3(a5)]. To test whether the islands may be formed from Se itself, we deposited Se onto a STO substrate at room temperature and then annealed the substrate, following a procedure similar to the one we used on single-layer FeSe/STO films. When Se is deposited on a clean STO substrate [Fig. 3(d1)], after a mild annealing at 150 °C for 0.5 *h*, a disorder structure is formed on the surface [Fig. 3(d2)] which can be related to the amorphous Se. After further annealing the sample at 150 °C for another 2 *h*, the amorphous Se totally disappears, and the sample recovers to the original clean STO surface, as shown in Fig. 3(d3). This indicates that Se capping is indeed easy to remove when it is deposited on the STO surface. This also rules out the possibility that the islands we observed on the single-layer FeSe/STO films are from pure Se.

Now we zoom in on the detailed structure of the islands. We first measured the height of the islands by scanning through them, like scans 2 and 3 in Fig. 4(a). For self-calibration, we also performed scan 1 in Fig. 4(a) going through a single-layer FeSe region crossing a step between the two terraces. The observed height difference is ~0.38 *nm* [Fig. 4(b1)], which is consistent with the step height on the initial STO substrate. The height of the island above the first FeSe layer on terrace 2 for scan 2 [Fig. 4(b2)] is 0.55 *nm*. It is also 0.55 *nm* for scan 3 above the single-layer FeSe on terrace 1 [Fig. 4(b3)]. We have measured the height of many islands, and it is always 0.55 *nm*. An atomic-resolution image was obtained on the surface of an island, as shown in Fig. 4(c), which facilitates the determination of its crystal structure. The in-plane lattice is tetragonal, and the lattice constant is ~0.38 *nm*, as determined by a line scan across the island surface [blue curve in Fig. 4(d)]. The in-plane tetragonal structure and the lattice constant are identical to those of the first FeSe layer [black curve in Fig. 4(d)]. This observation, together with the fact that the height of the islands is also identical to that of the single FeSe layer, strongly indicates that the islands represent second FeSe layers.

Now we discuss the origin of the formation of the second-layer FeSe islands. One possibility is that some first FeSe layers may migrate to form the second layer. We think it is unlikely for the following reasons: (1) The single-layer FeSe film is obtained by annealing at 530 °C; it is impossible for it to decompose at a very low temperature like 150 °C. (2) the direct STM measurements show that the first FeSe layer is intact during the annealing process. (3) This possibility cannot explain the morphology evolution during annealing and eventual recovery of the perfect single-layer FeSe [Fig. 3(a)]. A natural conclusion is that these FeSe islands are formed by the reaction of deposited Se with Fe during the annealing process. Since only Se is deposited on the sample surface, the formation of the extra second-layer FeSe indicates the existence of an appreciable amount of excess Fe in the single-layer FeSe/STO film. This observation provides a unique way to detect excess Fe and estimate its quantity from the island coverage of the second-layer FeSe after Se deposition and annealing.

One immediate question to ask is the source of this excess Fe in single-layer FeSe films. Is it possible that it comes from the leftover Fe of the second layer that decomposes during post-annealing at high temperature? It is true that in some as-grown FeSe films, we can observe the second-layer FeSe islands near the steps [Fig. 1(c)]. To check on this possibility, we have grown submonolayers with different thicknesses (0.45 and 0.75 ML) and 1.5-ML FeSe/STO films. We carried out Se capping and annealing processes similar to those done on the 1-ML FeSe/STO film [Fig. 3(a)]. Figure 5 shows STM images of the as-grown 0.45-, 0.75-, 1-, and 1.5-ML FeSe/STO films [Figs. 5(a1)–5(a4)] and the samples after Se deposition at room temperature followed by annealing [Figs. 5(b1)–5(b4)]. The as-grown 1.5-ML FeSe/STO film [Fig. 5(a4)] was annealed at a high temperature (520 °C) to first get a perfect 1.0 ML (see the Supplemental Material [18]). Then the 1.0-ML FeSe/STO film was capped by Se deposition at room temperature and annealed at 150 °C for 2 h. Figure 5(b4) shows the surface morphology of the sample where a large number of islands of the second FeSe layers are observed. The coverage of the islands is ~45%, indicating there is at least 45% excess Fe in the prepared 1.0-ML FeSe/STO film. For the as-grown 0.75-ML FeSe/STO film [Fig. 5(a2)], after Se deposition at room temperature and annealing at 250 °C for 2 *h*, a large number of islands are observed, as shown in Fig. 5(b2). The coverage of the islands, when normalized to the single-layer FeSe (ratio between the area of the islands and the area of 0.75-ML FeSe in the field of view), is about 23%. For the as-grown 0.45-ML FeSe/STO film [Fig. 5(a1)], the coverage of the islands after initial annealing at 250 °C for 2 *h* is ~20% [Fig. 5(b1)].

Figure 5(c) shows the variation of the normalized coverage of the second-layer FeSe islands with the thickness of the as- grown FeSe/STO films. The amount of excess Fe varies little for the as-grown 1-ML and submonolayer FeSe/STO films (20%–23%), while it is much higher in the as-grown 1.5-ML FeSe/STO film. Since no post-annealing process is involved in preparing the submonolayer FeSe/STO films, the detection of excess Fe above 20% in these samples indicates that there is already excess Fe present in the as-grown FeSe/STO films. When FeSe films decompose in the annealing process, it is plausible that some Fe will be left on the sample surface, as demonstrated by observing leftover Fe in the completely decomposed single-layer FeSe/STO film (see the Supplemental Material [18]). This helps us understand why the excess Fe in the 1-ML FeSe film (Fig. 5(d4) and the Supplemental Material [18]) is high (45%) because there are two contributions. The first source comes from the inherent excess Fe that is already present during the sample growth (~23%). The other source comes from the leftover Fe after the second-layer FeSe in the as-grown 1.5-ML FeSe/STO film [Fig. 5(a4)] decomposes during the annealing process to make a perfect 1.0-ML sample. This also helps us understand why the excess Fe can be very different for the 1-ML FeSe/STO films [23% in Fig. 3(a1), 36% in Fig. S1(b), and 45% in Fig. 5(b4)] because a perfect 1-ML FeSe/STO film is obtained by post-annealing as-grown FeSe/STO films with different thicknesses at a high temperature (~520 °C) to remove the second-layer FeSe. The total amount of excess Fe therefore depends on the sample growth condition (contribution of inherent excess Fe) and the initial thickness of the as-grown FeSe films (contribution of leftover excess Fe).

The next question that arises is where this excess Fe may be located in single-layer FeSe/STO films. There are three possible locations of the excess Fe in 1-ML FeSe/STO films: inside the FeSe layer, on the sample surface, and at the interface between the FeSe layer and the STO substrate. It is possible that some excess Fe may be accommodated inside the single FeSe layer; excess Fe can be located at the interstitial sites as detected in bulk $Fe_{1+\delta}(Se,Te)$ [22]. But it is unlikely that the single FeSe layer can accommodate a significant amount of excess Fe. It is found that superconductivity of bulk FeSe is very sensitive to a tiny amount of excess Fe and that less than 3% excess Fe can fully suppress the superconductivity [14]. For the 1-ML FeSe/STO film with excess Fe of ~23% (Fig. 1), since the amount of excess Fe is close to that of 0.45- and 0.75-ML FeSe/STO films [Fig. 5(c)], the excess Fe is mainly inherited during the growth process. The large amount of excess Fe is unlikely to be entirely on the surface of the FeSe layer. If Fe atoms are on the FeSe surface, they can be observed by STM [16]. Our atomic-resolution STM measurements [Fig. 1(f)], along with other STM measurements [1,23,24], find no indication of Fe atoms on top of the single-layer FeSe films. In this case, it is possible that the excess Fe lies at the interface between the STO substrate and single-layer FeSe. High-resolution scanning transmission electron microscopy (STEM) measurements on a similar superconducting single-layer FeSe/STO film reveal some unidentified extra atoms present at the interface [25]. Since we have revealed the presence of a large amount of excess Fe in the single-layer FeSe/STO film and it likely lies at the interface, further work needs to be done to determine whether the extra atoms observed by STEM at the interface may correspond to the excess Fe we have identified. When the amount of excess Fe is significantly increased in single-layer FeSe/STO films due to leftover Fe from decomposition of the second-layer FeSe, it is possible that some extra Fe may be present on the sample surface or even form Fe clusters (see the Supplemental Material [18]).

Our results indicate that there is a large amount of excess Fe above 20% present in the single-layer FeSe/STO films and the inherent part (~20%) is most likely located at the interface between the single-layer FeSe film and STO substrate. A distinct electronic structure and signature of superconductivity at about 65 $K$ were observed on 1-ML FeSe/STO films [2,3] that were prepared and annealed in a manner similar to those we used here. The discovery of a large amount of excess Fe (above 20%) in such superconducting single-layer FeSe/STO films with $T_C$ above 65 $K$ is quite surprising. This observation has important implications for understanding high-temperature superconductivity in single-layer FeSe/STO films. Since superconductivity in single-layer FeSe/STO films is realized by electron doping [2,3,23], the valence state of Fe and its role in the electron doping need to be evaluated [26]. Also, because Fe is usually magnetic, whether this excess Fe at the interface can form an ordered magnetic state and its influence on the magnetic state of the adjacent FeSe layer need to be further investigated [27–29]. The detection of excess Fe in single-layer FeSe/STO films, particularly the formation of the second-layer FeSe upon the deposition of Se on the surface and annealing, calls for reexamination of some results previously found for Se-protected FeSe/STO films. We point out that the formation of the second-layer FeSe does not affect our conclusions on the intrinsic electronic structures in superconducting single-layer FeSe/STO films [2], the existence of the N phase and S phase in the FeSe films [3,30], the observation of superconductivity at 65 $K$ [3], or the insulator-superconductor crossover in single-layer FeSe/STO films [31], although the origin of the N phase in single-layer FeSe/STO films [3] needs to be reconsidered [32].

## IV. CONCLUSION

In summary, by depositing Se onto the single-layer FeSe/STO films, we found the formation of second-layer FeSe islands on top of the first layer during the annealing process at a surprisingly low temperature (~150 °C) which is much lower than the usual growth temperature (~490 °C). This observation allows us to develop a reliable method to detect excess Fe and estimate its quantity in single-layer FeSe/STO films. We find that there is at least 20% excess Fe present that is surprisingly high for the superconducting single-layer FeSe/STO films. Our present work provides in- sights into understanding high-temperature superconductivity in single-layer FeSe/STO films. In establishing reasonable models for the FeSe/STO system, in addition to the STO substrate and single FeSe layer, the presence of a large quantity of excess Fe at the interface should be taken into account. Since magnetism is usually detrimental to superconductivity [14,33], the effect of this excess Fe on the superconductivity of single-layer FeSe/STO films deserves further investigation, and removal of the excess Fe may further enhance superconductivity.

**ACKNOWLEDGMENTS**

We thank Q. Xue, X. Ma, L. Wang, and F. Li for their help in growing our FeSe/SrTiO$_3$ films. We are thankful for financial support from the National Key Re-search and Development Program of China (Grants No. 2017YFA0302900 and No. 2016YFA0300300), the National Natural Science Foundation of China (Grants No. 11334010 and No. 11534007), the National Basic Research Program of China (Grants No. 2015CB921000 and No. 2015CB921300), and the Strategic Priority Research Program (B) of the Chinese Academy of Sciences with Grants No. XDB07020300 and No. XDPB01.

X.J.Z., Y.H., and Q.W. proposed and designed the research. Y.H. carried out the experiment with Y.X., Q.W., L.Z., S.H. and G.L.; Y.H., Q.W., L.Z., and X.J.Z. analyzed the data. X.J.Z. and Y.H. wrote the paper with Q.W. and L.Z. All authors discussed the results and commented on the manuscript. Y.H. and Y.X. contributed equally to this work.

The authors declare no competing financial interests.

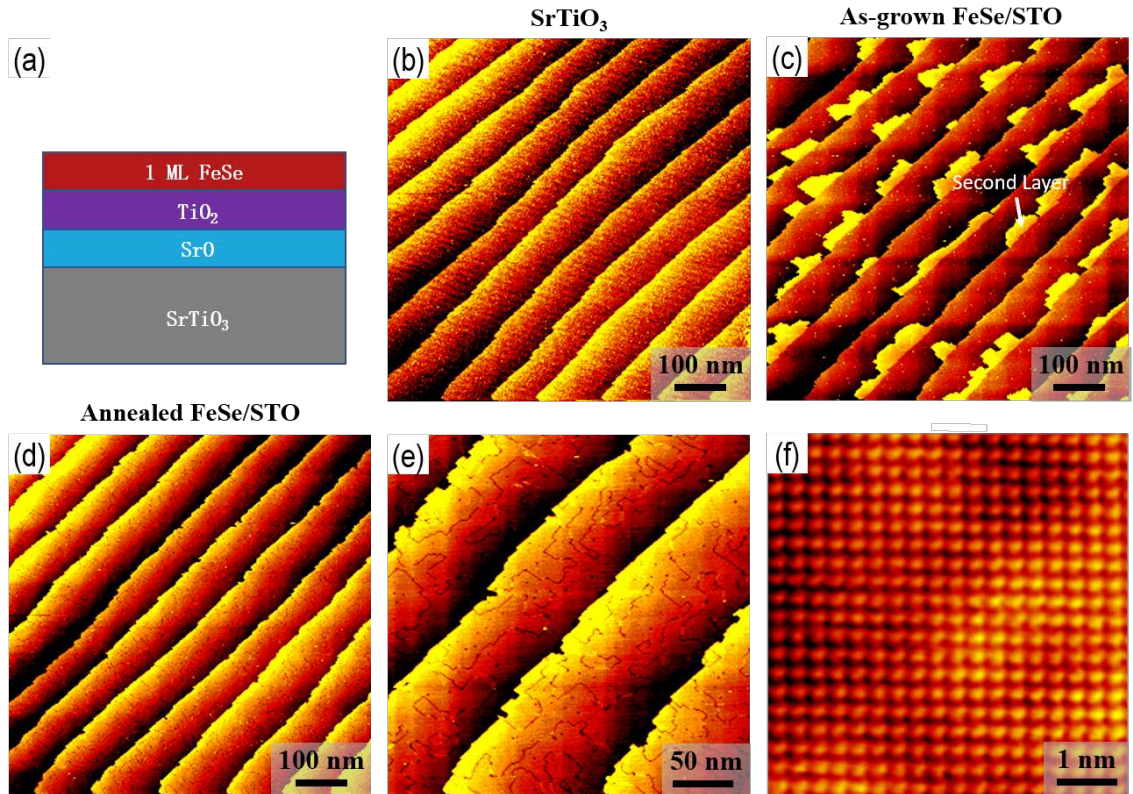

**FIG. 1. Preparation of the single-layer FeSe/STO film.** (a) Schematic of the FeSe film grown on the STO substrate. (b) STM topography (700 × 700 $nm^2$) of the atomically flat STO (001) surface. (c) The as-grown 1-ML FeSe film on the STO (001) substrate with some second-layer FeSe patches (yellow) as marked by the white arrow. (d) A perfect 1-ML FeSe/STO film after annealing the film in (c) at 530 °C for 4 *h* and (e) its zoomed-in image and (f) the atomically resolved image.

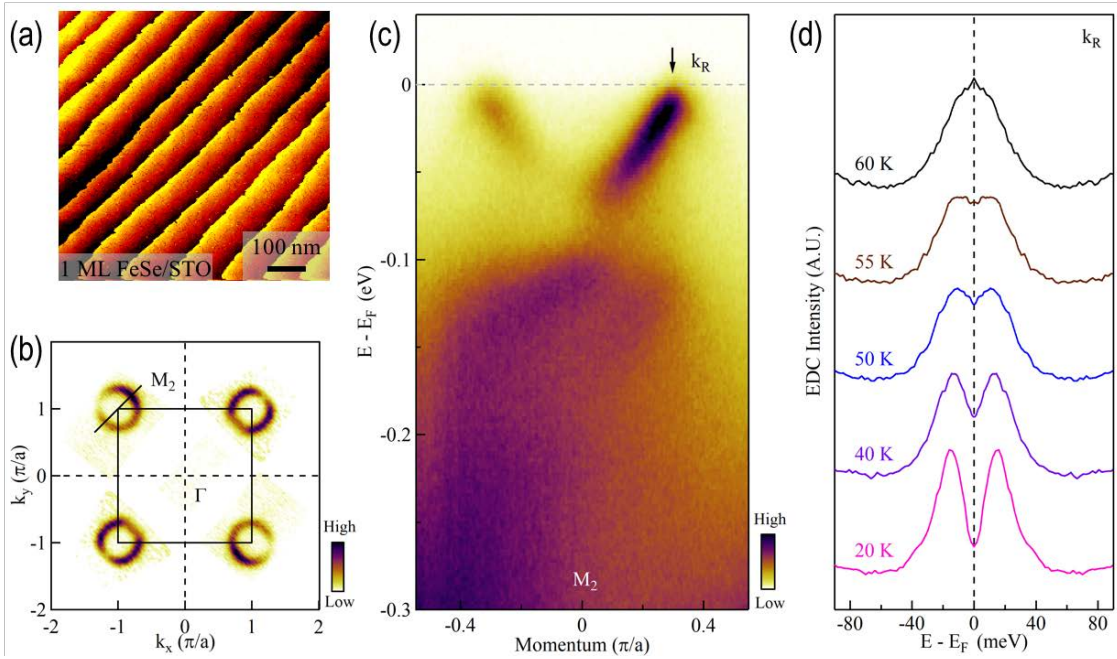

**FIG. 2. High-temperature superconductivity in the single-layer FeSe/STO film.** (a) STM topography (700 × 700 $nm^2$) of a perfect 1-ML FeSe/STO film after annealing at 530 °C for 4 *h*. (b) Fermi surface of the annealed single-layer FeSe/STO film in (a), measured at 20 *K*. The carrier concentration is ~0.116 electrons/Fe atom as determined from the area of the electron pockets near M. (c) Band structure of the single-layer FeSe/STO film along a momentum cut near $M_2$ as marked in (b). (d) Symmetrized photoemission spectra (energy distribution curves, EDCs) at the Fermi momentum $k_R$ marked in (c) measured at different temperatures. The superconducting gap size is ~15 *meV* at 20 *K* and it closes around ~60 *K*.

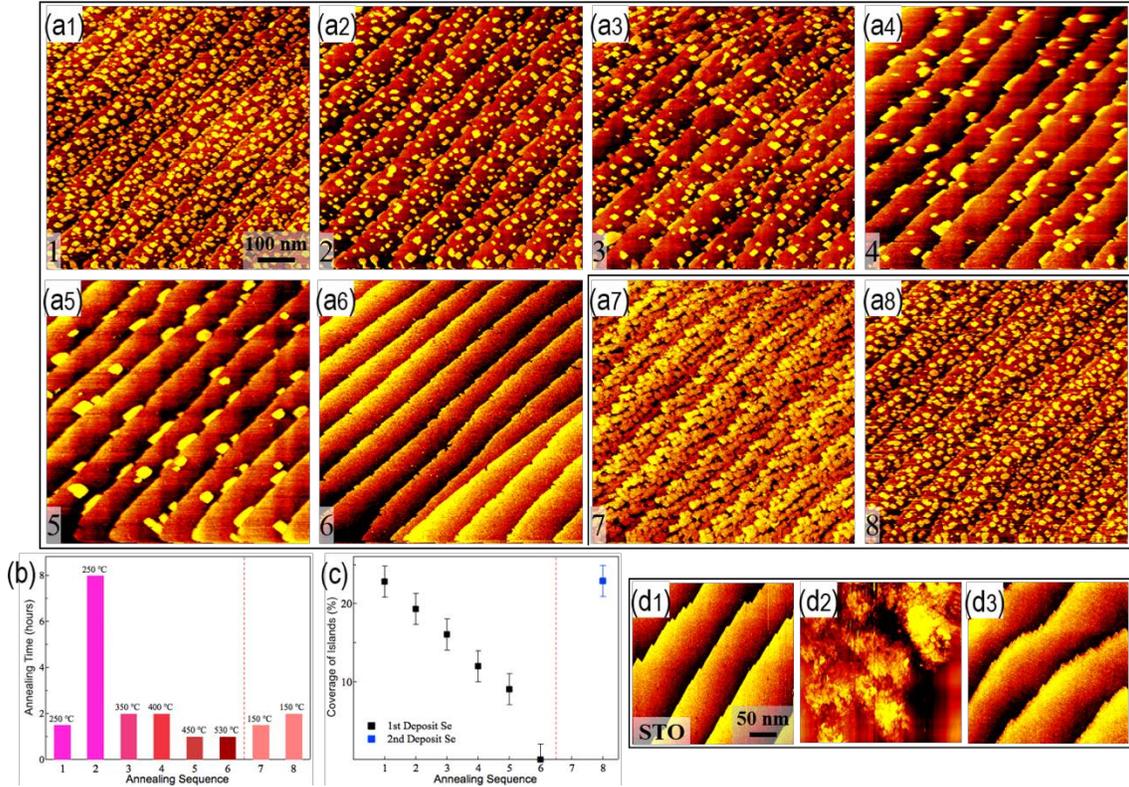

**FIG. 3. Se deposition on the single-layer FeSe/STO film and evolution of the surface morphology with annealing.** (a1)–(a8) STM topography (700 × 700 $nm^2$) after each annealing sequence. In the first five sequences [(a1)–(a5)], the islands grow in size, and their total area decreases with annealing. The islands completely disappear after annealing sequence 6 [(a6)]. Then Se is deposited again on the single-layer FeSe film [(a6)]. (a7) and (a8) show STM images after annealing sequences 7 and 8, respectively. (b) The annealing sequence after depositing Se onto the 1-ML FeSe thin film. For convenience, we use 1–8 to denote eight consecutive annealing conditions. (c) The coverage of islands on the 1-ML FeSe film after each annealing sequence. Images of (d1) an atomically flat STO surface and the surface morphology after deposition of Se on a STO surface followed by (d2) annealing at 150 °C for 0.5 *h* and (d3) another annealing at 150 °C for 2 *h*.

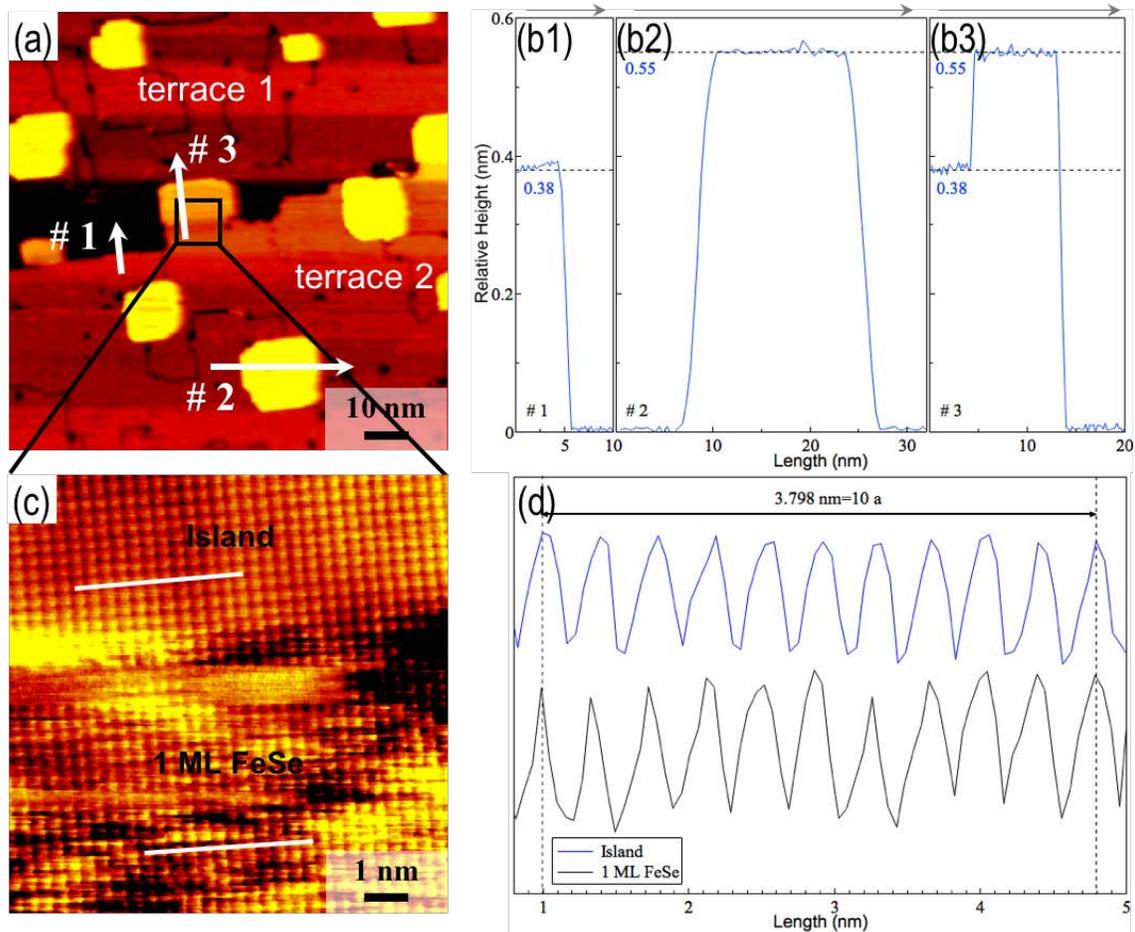

**FIG. 4. STM characterization of the islands on the single-layer FeSe/STO film.** (a) STM image (100 × 100 $nm^2$) after the third annealing sequence [Fig. 3(a3)]. It consists of two single-layer FeSe terraces and several islands (in yellow) on the top of the surface. (b1)–(b3) Line profiles along cuts 1, 2, and 3, respectively, as marked by white lines in (a). (c) Expanded view of the region marked by a black square in (a). It shows an atomically resolved STM image (10 × 10 $nm^2$) consisting of both the single-layer FeSe [terrace 2 in (a)] and an island. (d) Line profiles along the two cuts marked by the white lines in (c). The upper blue curve represents the profile along the cut on the island surface, while the lower black curve is for the cut on the surface of single-layer FeSe.

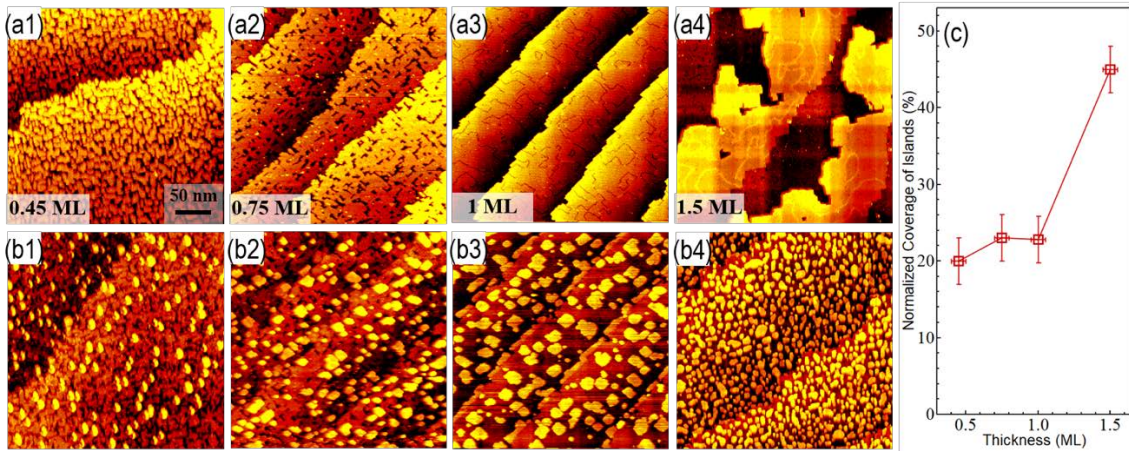

**FIG. 5. Detection of excess Fe in FeSe/STO films with different thickness.** STM images (300 × 300 $nm^2$) of (a1) the as-grown 0.45-ML FeSe/STO film, (a2) 0.75-ML FeSe/STO film, (a3) 1-ML FeSe/STO film, and (a4) 1.5-ML FeSe/STO film. (b1) STM image of the 0.45-ML FeSe/STO film after Se deposition and annealing at 250 °C for 2 *h*. (b2) STM image of the 0.75-ML FeSe/STO film after Se deposition and annealing at 250 °C for 2 *h*. (b3) STM image of the 1-ML FeSe/STO film after the as-grown FeSe film was annealed at 530 °C for 4 *h*, followed by Se deposition and annealing at 150 °C for 2 *h*. (b4) STM image of the 1-ML FeSe/STO film after the as-grown 1.5-ML FeSe film was annealed at 530 °C for 9 *h*, followed by Se deposition and annealing at 150 °C for 2 *h*. (c) Normalized coverage of the second-layer FeSe islands for FeSe/STO films with different initial thicknesses.

**Supplementary Materials**

# Identification of a Large Amount of Excess Fe in Superconducting Single-Layer FeSe/SrTiO$_3$ Films

Yong Hu[1,2,#], Yu Xu[1,2,#], Qingyan Wang[1,*], Lin Zhao[1], Shaolong He[1], Jianwei Huang[1,2], Cong Li[1,2], Guodong Liu[1] and X. J. Zhou[1,2,3,*]

[1]National Lab for Superconductivity, Beijing National Laboratory for Condensed Matter Physics, Institute of Physics, Chinese Academy of Sciences, Beijing 100190, China

[2]University of Chinese Academy of Sciences, Beijing 100049, China 3Collaborative Innovation Center of Quantum Matter, Beijing 100871, China

[#]These authors contributed equally to the present work.

[*]Corresponding author: qingyanwang@iphy.ac.cn, XJZhou@iphy.ac.cn

## 1. Reversible process of Se deposition and annealing on a single-layer FeSe/SrTiO$_3$ film

We prepared another single-layer FeSe/STO film (Fig. S1) and repeated Se-deposition and annealing procedures for several times. For convenience, we use 1-6 to denote the six procedures. For each procedure, after the perfect single-layer FeSe film is obtained from annealing at a high temperature (450-520 °C) for 2-3 hours, Se is deposited on the surface at room temperature and annealing at 150 °C for 2 hours. For the first Se deposition on the perfect 1 ML FeSe/STO film, the coverage of the islands is about 36% (left-most panel in Fig. S1) which is different from that of the 1 ML FeSe/STO films in Fig. 3(a1) (23%). This is because the initial thickness of the as-prepared FeSe/STO films is different for these two samples.

In the first 5 procedures, the coverage of islands for each sequence is similar, although the coverage of the single-layer FeSe film shows a slight decrease due to sample decomposition. These results indicate that the above Se deposition and annealing processes are highly reproducible. In the sixth sequence, we annealed the sample at 600 °C for 1 hour, the single-layer FeSe almost entirely decomposes and some yellow regions are formed that may be Fe clusters (Fig. S1(c)). Se is deposited again on the sample at room temperature and the sample is then annealed at 150 °C for 2 hours. It still has some clusters on the sample surface (Fig. S1(d)) without going back to the single-layer FeSe film.

## 2. Preparation of 1.0 ML FeSe/SrTiO3 film by annealing the as-grown 1.5 ML FeSe/SrTiO$_3$ film

Figure S2 shows the preparation procedure to obtain a perfect 1.0 ML FeSe/STO film from annealing the as-grown 1.5 ML FeSe/STO film. The STO substrate we used here has an atomically flat terraces (Fig. S2(a)). The as-grown 1.5 ML FeSe film on the STO (001) substrate has 0.5 ML second-layer FeSe (yellow region marked in Fig. S2(b)). The extra second-layer FeSe decreases during post-annealing the as-grown FeSe film at 520 °C (Fig. S2(c) and (d)) and completely disappears to get a perfect 1.0 ML FeSe/STO film (Fig. S2(e)). Then the 1.0 ML FeSe/STO film was covered by Se at room temperature and annealed at 150 °C for 2 hours. Figure S2(f) shows the surface morphology of the sample where a large amount of islands of the second FeSe layers are observed. The coverage of the islands is ~45% indicating there is at least 45% excess Fe in the prepared 1.0 ML FeSe/STO film (Fig. S2(e)).

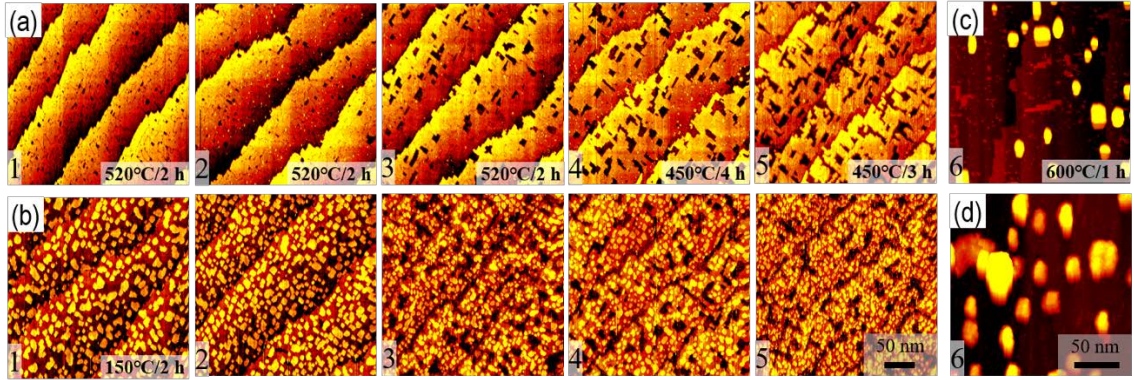

**FIG.S1: Reversible process of Se deposition and annealing on a single-layer FeSe/STO film.** (a) STM images (300 × 300 $nm^2$) of the nearly perfect single-layer FeSe/STO film after high temperature annealing. The annealing conditions for each sequence are labeled. After the perfect single-layer FeSe film is obtained, Se is deposited on the sample at room temperature. The sample is then annealed at 150 °C for 2 hours and the corresponding surface morphologies are shown in (b). (c) STM image of the FeSe film after annealing at 600 °C for 1 hour. The single-layer FeSe almost entirely decomposes and some clusters (yellow regions) are observed on the sample surface. Then Se is deposited again on the sample (c) at room temperature and the sample is annealed at 150 °C for 2 hours. The surface morphology is shown in (d).

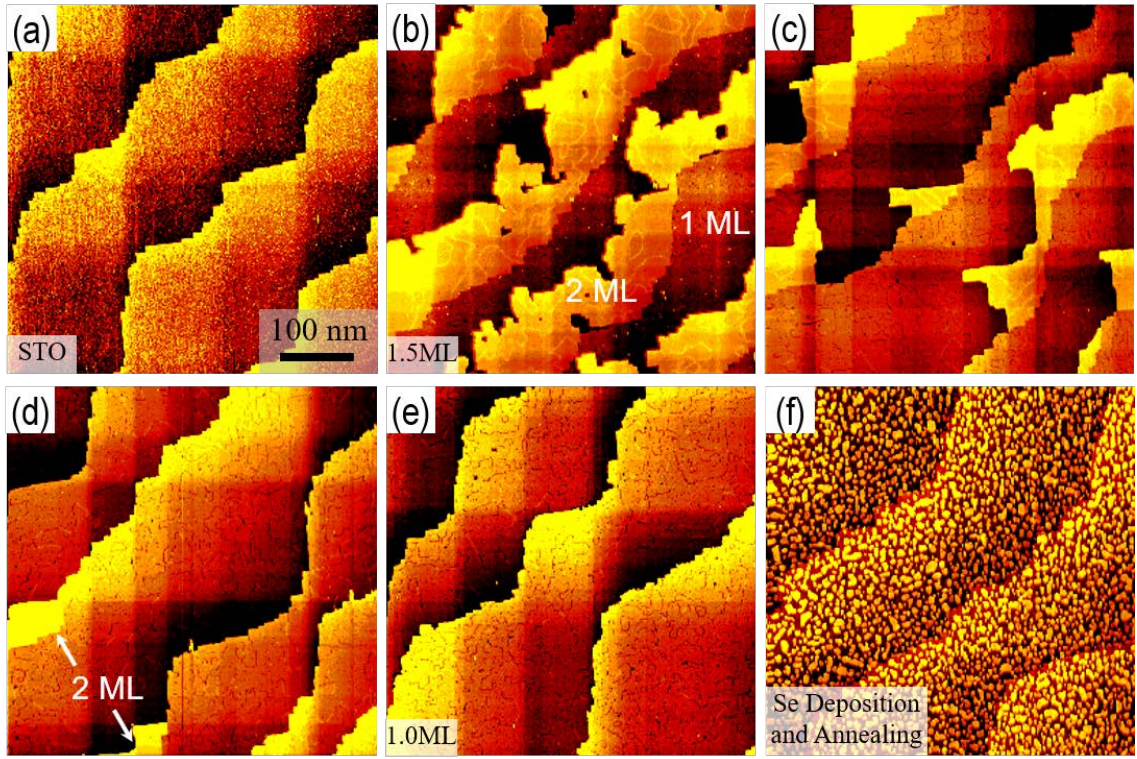

**FIG.S2: Preparation of 1.0 ML FeSe/STO film by annealing the as-grown 1.5 ML FeSe/STO film.** (a) STM topography (500 × 500 $nm^2$) of the atomically flat STO (001) surface. (b) The as-grown 1.5 ML FeSe film on the STO (001) substrate with 0.5 ML second-layer FeSe which are yellow in color as labeled in the figure. (c)-(e) show surface morphologies after annealing the as-grown 1.5 ML FeSe/STO film at 520 °C for 5 hours (c), for another 2 hours (d) and for another 2 hours (e). A perfect 1.0 ML FeSe/STO film is obtained (e). Then Se is deposited on the film (e) at room temperature and the sample is annealed at 150 °C for 2 hours. The surface morphology of the sample is shown (f).